\documentclass[aps,prl,reprint,amsmath,superscriptaddress,amssymb,showpacs]{revtex4-1}
\usepackage[english]{babel}
\usepackage{graphicx}
\usepackage{dcolumn}
\usepackage{bm}
\usepackage[utf8]{inputenc}
\usepackage{setspace}
\usepackage{multirow}
\usepackage{color}

\newcommand{\um}[1]{\,\rm{#1}}

\makeatletter
\renewcommand{\fnum@figure}{FIG. \thefigure}
\makeatother

\newcommand{\beq}{\begin{equation}}
\newcommand{\eeq}{\end{equation}}
\newcommand{\be}{\begin{equation}}
\newcommand{\ee}{\end{equation}}
\newcommand{\beqa}{\begin{eqnarray}}
\newcommand{\eeqa}{\end{eqnarray}}
\newcommand{\bea}{\begin{align}}
\newcommand{\eea}{\end{align}}

\newcommand{\req}[1]{Eq.~(\ref{#1})}

\newcommand{\rref}[1]{(\ref{#1})}

\begin{document}

\title{Transient superconductivity without superconductivity}
\author{Giuliano Chiriacò}
\affiliation{Department of Physics, Columbia University, New York, NY 10027}
\author{Andrew J. Millis}
\affiliation{Department of Physics, Columbia University, New York, NY 10027}
\affiliation{Center for Computational Quantum Physics, The Flatiron Institute, New York, NY 10010}
\author{Igor L. Aleiner}
\affiliation{Department of Physics, Columbia University, New York, NY 10027}

\date{\today}

\begin{abstract}
Recent experiments on K$_3$C$_{60}$ and layered copper-oxide materials have reported substantial changes in the optical response following application of an intense THz pulse. These data have been interpreted as the stimulation of a transient superconducting state even at temperatures well above the equilibrium transition temperature. We propose an alternative phenomenology based on the assumption that the pulse creates a non-superconducting, though non-equilibrium situation in which the linear response conductivity is negative. The negative conductivity implies that the spatially uniform pre-pulse state is unstable and evolves to a new state with a spontaneous electric polarization. This state exhibits coupled oscillations of entropy and electric charge whose coupling to incident probe radiation modifies the reflectivity, leading to an apparently superconducting-like response that resembles the data. Dependencies of the reflectivity on polarization and angle of incidence of the probe are predicted and other experimental consequences are discussed.
\end{abstract}


\maketitle

There has been substantial interest in the use of intense radiation
fields to drive materials into non-equilibrium states
\cite{Av:Tay}. Particular excitement has been generated by
 reports \cite{K3:C60,LBCO1,YBCO} of dramatic changes in
 the electromagnetic response
of K$_3$C$_{60}$ and layered copper-oxide
  materials after their exposure to intense THz radiation. The key
features of the data are: i) before the application of the pump
pulse, the material is in the normal (unbroken symmetry) state; ii)
after photo-excitation of the material by the pump, the reflectivity
$R(\omega)$ is measured as a function of the frequency $\omega$ of a
probe field; iii) for some time after the pump excitation, $R(\omega)$
is found to be substantially enhanced at low frequency, see the insets in Fig. 1. This enhancement has been interpreted in
terms of the creation, by the pulse, of a superconducting (SC) state.

Theories proposed to date \cite{DMK:AJM,Deml1,Georges,Oka,Fab,Lem,Kim}
are all based on the premise that the pump pulse changes the
interactions and/or structure in a way that enables a transition to a
broken symmetry SC state at a temperature much higher
than that of the equilibrium transition. In this
work we point out that the data do not require this
interpretation; instead the observations can be understood within a
general phenomenology that does not involve SC.

The essence of our model is: i) we argue on general grounds that a
non-equilibrium system can exhibit a negative linear response
conductivity; ii) in this case the spatially homogeneous state is
unstable and evolves to a new state characterized by domains of
constant electric field bounded by sheets of charge,
Fig. 2a; iii) in the experimentally relevant situation where the
non-equilibrium state is produced by a  pulse and thereafter
evolves with a conserved energy, we show that the system sustains
collective modes strongly coupled to incident radiation,
leading to the reflectivity curves  shown in Fig. 1.
\begin{figure}[t]\label{F1}
\centering
\includegraphics[width=0.48\columnwidth]{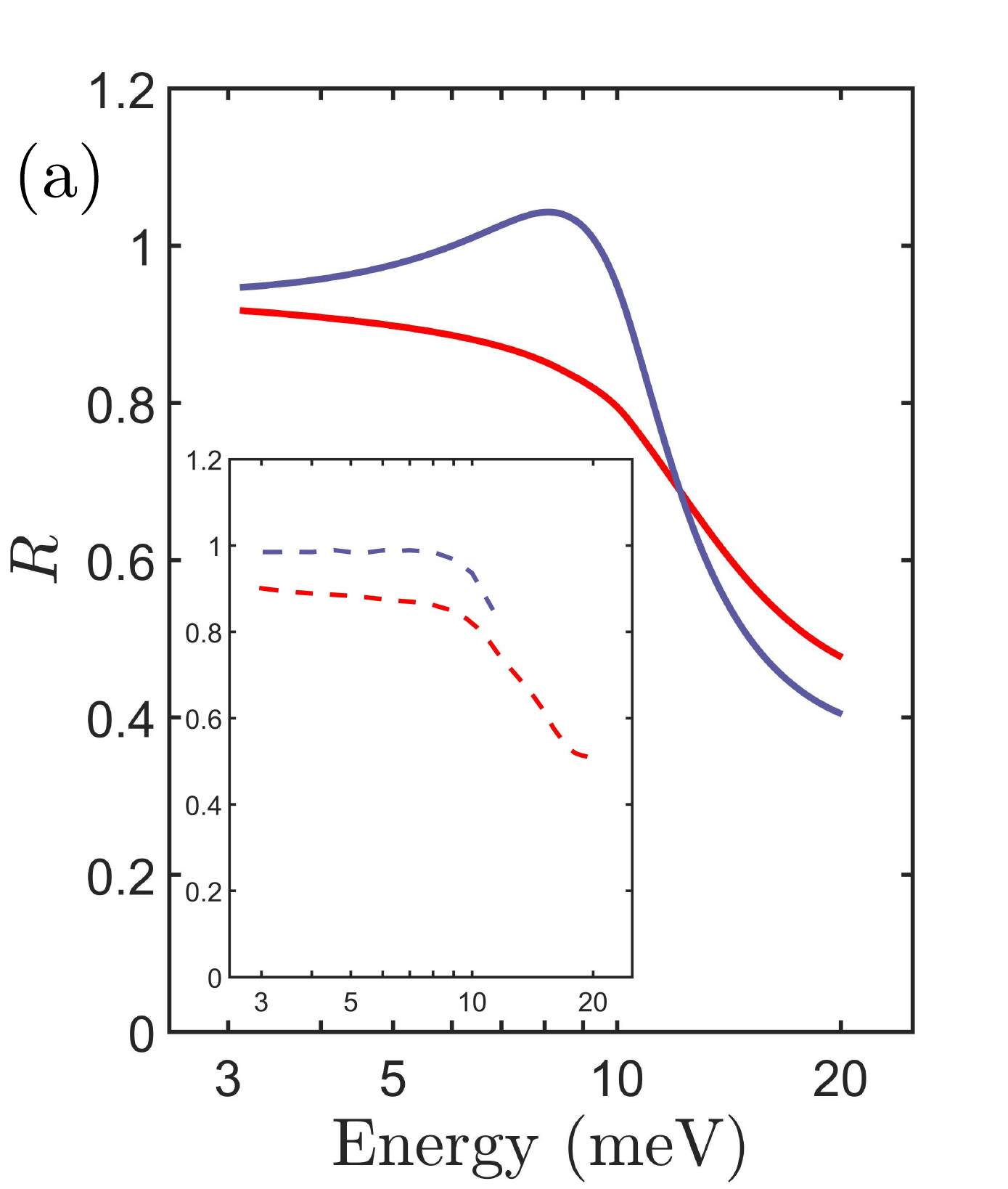}\,\,\,\,
\includegraphics[width=0.48\columnwidth]{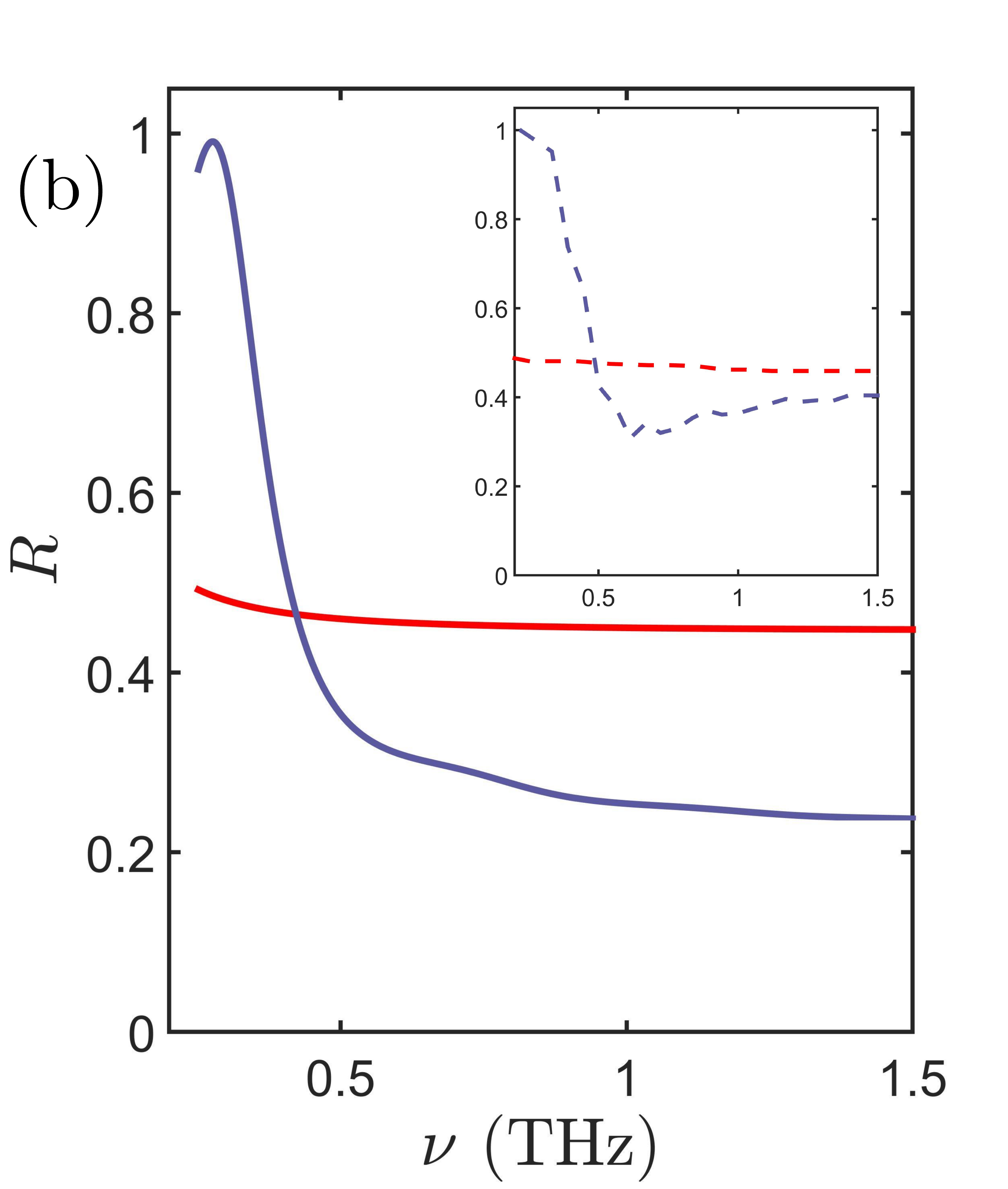}
\caption{\scriptsize{(Color online) Calculated (solid lines) and experimental (insets, dashed lines) reflectivities for K$_3$C$_{60}$ (a) and La$_{2-x}$Ba$_x$CuO$_4$ (LBCO) (b) for equilibrium (red) and non-equilibrium (blue) situations. In (a) the  data are taken from Fig. 2a of Ref. \cite{K3:C60} and the calculations are done as described in the text for angle of incidence $\theta=45^{\circ}$ using parameters $\omega_E=110\um{THz}$, $\gamma=3.2\um{THz}$, $l_0=600\um{\mbox{\AA}}$, $\kappa =3\um{cm^2s^{-1}}$. In (b) the  data are taken from Fig. 2b2 of Ref. \cite{LBCO1} and the solid curves are calculated for $\theta=45^{\circ}$ using $\omega_E=1200\um{THz}$, $\gamma=0.6\um{THz}$, $l_0=4500\um{\mbox{\AA}}$, $\kappa =0.2\um{cm^2s^{-1}}$. The anisotropy of LBCO was not considered. The non-equilibrium data of Refs. \cite{K3:C60,LBCO1} are processed from raw data and report $R(\omega)$ as if the thickness of the non-equilibrium layer were infinite, thus magnifying the non-equilibrium effects on $R$; a direct quantitative comparison with our calculations is not possible, but the resemblance of the curves is very reasonable.}}
\end{figure}

{\em i)} Consider the system out of equilibrium. The sample occupies the half space $z>0$. Pump radiation incident from $z<0$ creates a non-equilibrium situation, which we assume relaxes rapidly to a quasi-steady non-equilibrium state; in the simplest case this state is characterized by one  parameter, $\zeta(\vec r,t)$, which relaxes slowly to its equilibrium value $\zeta=0$. The precise microscopic description of $\zeta$ is not important here. For $\zeta\neq 0$, entropy
density, $S$, is produced; we describe this production by a generation
function $G_0(\zeta)$ with $G_0(\zeta\neq 0)>0$. Electric fields $E$
and currents $j$ produce entropy via the Joule heating
term, $jE$, leading to ($T$ is a pseudo-temperature defined later)
\begin{equation}\label{Sprod2}
T \partial_t S=\sigma\vec{E}^2+G_0(\zeta)=\rho\vec j^2+G_0(\zeta).
\end{equation}
Here, the conductivity (resistivity) $\sigma\,(\rho)$ depends on $\zeta,T$.

The second law of thermodynamics requires $\frac{dS}{dt}\geq0$. At
equilibrium $G_0(\zeta=0)=0$, implying $\rho\geq0$. This means that in
a system which is superconducting ($\rho=0$ for a range of $T$),
$\rho(T)$ cannot be an analytic function of $T$: in other words, the
onset of a superconducting state is necessarily via a phase transition
(gauge symmetry breaking). However, in non-equilibrium,
$G_0>0$ so $\rho$ or $\sigma$ can cross zero without any
non-analyticity. Indeed, calculations have found negative
conductivities in several models of continuously driven systems
\cite{Zakh,Rizhi,*Rizhi2,IA:AJM,Sach,Dyak,*Dyak2,Vavilov}
and other models with similar properties may exist
\footnote{Existing calculations identify two origins of the negative conductivity: the photovoltaic effects  \cite{Rizhi,*Rizhi2,Sach,Dyak,*Dyak2} on the impurity collision process, and the distribution function effects \cite{Vavilov}. In slowly relaxing non-driven systems only the latter effects are relevant.}.

{\em ii)}
A state with $\vec E=0$ and $\sigma<0$ is unstable towards
formation of domains of electric field, see Fig. 2a.
To see this, we combine the continuity and Poisson equations
\be
\nabla\cdot\vec j_D=0;\quad \vec j_D=\vec
j+(4\pi )^{-1}\partial_t{\vec D}; \quad \nabla \times \vec{E}=0,
\label{1Maxwell}
\ee
with the constitutive equation $\vec{j}=\sigma \vec{E}$. Here
$\vec D=\epsilon_r\vec E$ is the electric displacement and
$\epsilon_r$ is the electric permittivity (for simplicity we treat $\epsilon_r,\sigma$ as isotropic). If $\sigma <0$, the $E=0$ solution is unstable: small fluctuations
in $E$ ({\em i.e.} charge) grow exponentially with time.
Then, the non-linear dependence of the
current on the electric field becomes important. In particular, at
some finite value of the electric field $E=E^*(\zeta)$ the Joule heating
vanishes again ($\vec{E}^*\vec{j}(E^*,\zeta)\to 0$), see Fig 2b and Refs.~\cite{IA:AJM,Dyak,*Dyak2}, implying  the formation of a
state characterized by domains of electric field $\sim E^*$ bounded by thin
sheets of electric charge.  The thickness of the sheets is determined
by   microscopic scales, and is not important for the physics we consider.
As usual for non-linear equations, a
multiplicity of possible domain structures may occur. Their
detailed analysis is a formidable but often unnecessary task and they were studied extensively in several works \cite{*[{See e.g. }][]
scholl,*Volk,*Gunn}.
\begin{figure}[t]
\centering
\includegraphics[width=0.82\columnwidth]{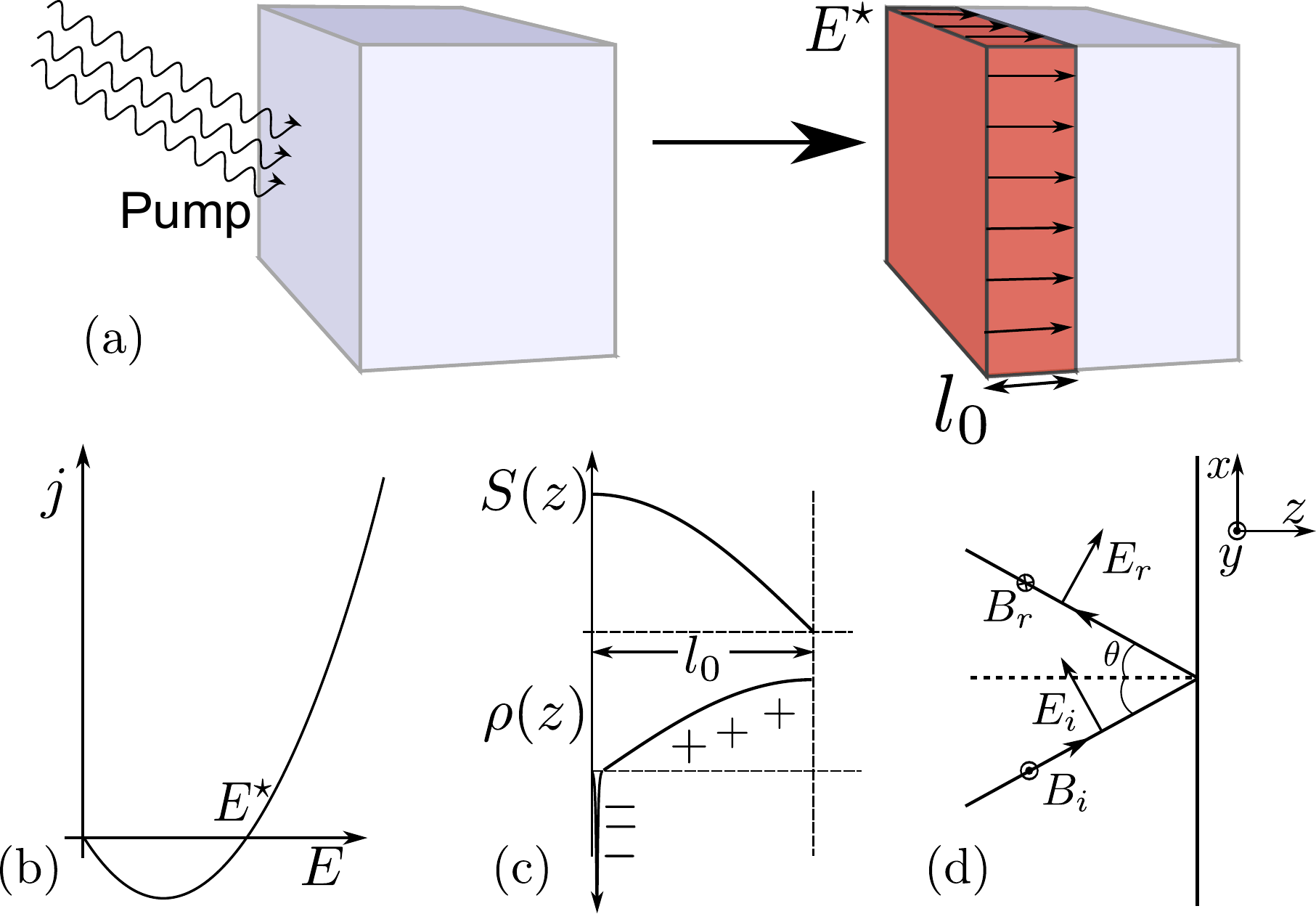}
\caption{\scriptsize{(Color online) (a) Sketch of incident pump pulse leading to  spontaneous polarization in the active layer.
(b) Sketch of the $j$-$E$ characteristic for a $\sigma<0$ state.
(c) Entropy and charge density profiles inside the active layer; notice the thin surface charge accumulation on the external boundary of the layer.
(d) Incident and reflected probe waves at an angle $\theta$; the reflection occurs in the $x-z$ plane and the magnetic field is along the $y$ axis for TM polarization.}}
\label{F2}
\end{figure}

{\em iii)} We study the physical consequences of  domain formation under the
main assumption that the non-equilibrium effects are
strong enough to have $\sigma(E=0)<0$ in some region near the sample
surface, leading to the formation of a spontaneous polarization
$E^{*}(z)$ in an {\em active layer} $0<z<l_0$ [{shaded region (red on-line) in the right portion} of Fig. 2a]. We
assume that the depth, $l_0$, of this active layer and the spontaneous
polarization change slowly with $\zeta$ as the system relaxes to
equilibrium, and {that  $E^*$ is determined by the dynamics of $\zeta(z,t)$, apart from the
small fluctuations considered below}.

After the pump is switched off, the microscopic degrees of freedom
rapidly relax to their quasi-equilibrium values; in particular, the
electric field relaxes to $E^*[\zeta(z,t)]$. Because the system
is no longer driven, the total energy is conserved, so the state is
characterized by three slowly evolving variables: the
parameter $\zeta$, the energy density $\varepsilon(\vec r,t)$,
and the electric field $E$ [connected to the charge and current densities by \req{1Maxwell}].
The entropy density $S$ is related to these dynamical variables by the equation of state
$S(\varepsilon,\vec{D},\zeta)$.

Conservation of energy means the
energy density (which includes the electric field energy)
evolves only via the energy current $\vec j_{\varepsilon}$
\begin{equation}\label{dte}
{\partial_t\varepsilon}+\vec{\nabla}\cdot\vec j_{\varepsilon}=0.
\end{equation}
Let us note in passing that the Joule heating increases the internal
energy of the electron system but decreases the energy of the electric
field so that it cancels from \rref{dte}.

The time evolution of $\zeta$ depends on $\varepsilon$ as a parameter
(since for homogeneous systems $\varepsilon$ is an integral of motion) and on $\zeta$ itself
\footnote{Note that $\zeta$ can always be chosen in such a way that the
right hand sides of Eqs. \eqref{dtf} and \eqref{dtS} do not depend on $S$.},
\be
\label{dtf} {\partial_t \zeta}=-I(\zeta,\varepsilon);\qquad I(\zeta=0,\varepsilon)=0.
\ee

For the  third dynamical equation we choose the entropy $S$ within the active layer as the independent variable with $\vec D$  determined from the equation of state:
\be
\label{dE}
\left(4\pi\right)^{-1}{\vec E\cdot d\vec D}=
d\varepsilon-TdS+T\partial_\zeta S d\zeta.
\ee
This choice enables us to use the conservation of energy \rref{dte} effectively. The entropy evolution can be written as
\be
\label{dtS}
T{\partial_t S}=G(\zeta,\varepsilon)>0;
\qquad T^{-1}=\left({\partial S}/{\partial\varepsilon}\right)_{\zeta,\vec{D}}.
\ee
Entropy generation arises both from  Joule heating and from the entropy produced by the relaxation of $\zeta$. The two effects cannot be separated in the non-equilibrium regime we consider and that is why they are joined in one kinetic term $G$.
However, the ratio $G/I=-T(\partial S/\partial \zeta)_{\varepsilon,\vec{D}}$ is determined directly by the state function; it is analogous to a
thermodynamic quantity and does not depend on the kinetic coefficients.

To complete the system of equations we observe that in the lowest order of the gradient expansion
\begin{equation}\label{je}
\vec j_{\varepsilon}=-\kappa T\vec{\nabla} S,
\end{equation}
where $\kappa $ is the thermal diffusion coefficient (related to
the thermal conductivity via the specific heat) \footnote{Strictly
  speaking Eq. \eqref{je} implies the presence of the divergence of
  the entropy current $\vec j_S=-\kappa \vec{\nabla} S$ in the LHS and
  of terms $\sim(\vec{\nabla}S)^2$ contributing to entropy production
  in the RHS of Eq. \eqref{dtS}. Both these terms are negligible in
  the studied regime $\omega\gg \kappa /l_0^2$.}.  The contribution of
the particle current to the energy current can be neglected provided
that all the relevant linear scale are much larger than the screening
radius.

{Equations \rref{dte}--\rref{je} provide a complete description of
the dynamics of the system in the non-relativistic limit (speed of light
$c\to \infty$) and in the absence of incident radiation.} 
It is noteworthy that \req{dE} shows that in the
 situation considered here, $\vec{E}\simeq \vec{E^*}$,
fluctuations of energy and entropy are linearly coupled to the electric field, in contrast to equilibrium
where the linear coupling is only via Seebeck and Peltier effects which involves only
spatial derivatives of the electric field.

For an isolated system $\left(\varepsilon_0,S_0,\zeta_0\right)_{(z,t)}$
slowly evolve according to Eqs. \eqref{dte}--\eqref{je}. Let us
consider small deviations around this evolving state: we
write $\varepsilon(\vec r,t)=\varepsilon_0(\vec r,t)+\delta\varepsilon$ {\it etc}
and linearize Eqs. \eqref{dte}--\eqref{je}, obtaining
\begin{equation}\label{L}
\hat{L}\begin{pmatrix} \delta \zeta\\ \delta\varepsilon \\ T\delta S\end{pmatrix}=0;\quad
\hat L=\begin{pmatrix}
\frac{\partial}{\partial t}+\frac{\partial I}{\partial \zeta} & \frac{\partial I}{\partial\varepsilon} & 0 \\
0 & \frac{\partial}{\partial t} & -\kappa \frac{\partial^2}{\partial z^2}\\
-\frac{\partial G}{\partial \zeta} & -\frac{\partial G}{\partial\varepsilon} & \frac{\partial}{\partial t}\end{pmatrix},
\end{equation}
with boundary conditions
\begin{equation}\label{BC}
\left.\partial_zT\delta S\right|_{z=0}=0,\qquad \delta S|_{z=l_0}=0.
\end{equation}
The first boundary condition {says entropy does not flow into the vacuum}; the second one states that any excitation reaching the internal boundary of the active layer is removed into the bulk \footnote{It can be checked that this assumption is justified for $\omega\gg \kappa /l_0^2$, which corresponds to a very small penetration length of the entropy current inside the bulk}.
Notice that $\delta\varepsilon$ can be discontinuous {at boundaries} due to  charge accumulation.

The coefficients in $\hat L$ depend slowly on time, justifying the use
of a quasi-stationary approximation for the response to rapidly varying
perturbations (perturbation frequency
$\omega\gg\frac{\partial I}{\partial \zeta}$). For simplicity we also
assume that all coefficients of $\hat L$ in Eq.~\eqref{L} do not
depend on $z$ within the layer $0<z<l_0$ (lifting this assumption leads to
unimportant changes in numerical coefficients). We seek solutions of the form
\be\label{adiab}
\delta \zeta,\delta\varepsilon,T\delta S\sim e^{-i\int^tdt_1\omega(t_1)}\cos(kz).
\ee

We define
\begin{equation}\label{param}
  \omega_E\equiv\partial_\varepsilon G;\qquad
  \gamma\equiv \partial_\varepsilon I
  \left({\partial_\varepsilon G}\right)^{-1}\partial_\zeta G,
\end{equation}
and substitute Eq. \eqref{adiab} into Eq. \eqref{L}, finding
\begin{equation}\label{k2}
k(\omega)={\omega}/{\sqrt{\kappa \omega_E(1-i\gamma/\omega)}}.
\end{equation}

From the second boundary condition in
Eq. \eqref{BC} $k^jl_0=\pi\left(j+\frac12\right)$; the lowest eigenfrequency is then:
\begin{equation}\label{w0}
2\omega_0\approx \sqrt{\pi^2\omega_E\kappa/ {l_0^2}}-i{\gamma}.
\end{equation}
The frequency $\omega_0$ will determine the scale of the non-equilibrium anomaly in the reflectivity and depends on time through $\omega_E$ and $l_0$. Equation \eqref{w0} shows that the active layer sustains underdamped fluctuations, originating from the combination of plasmonic charge
dynamics, the relaxation of $\zeta$ and slow fluctuations of the
energy. The coupling between the charge and energy/entropy
fluctuations is large because charge fluctuations produce electric
fields which contribute to the energy density, while even small
changes in $\varepsilon$ cause large changes in the entropy
production. Other oscillations involving such quantities such as the spin
density are possible, but their coupling to
the entropy fluctuations will be much weaker, as the interactions with
the corresponding densities are local.

We now turn to  the reflectivity. Coupling the collective mode \rref{w0} to electromagnetic wave requires replacing \req{1Maxwell} with the complete set of Maxwell equations
\begin{equation}\label{MEq}
c\vec{\nabla}\times\vec B={4\pi}\vec j_D;\qquad c\vec{\nabla}\times\vec E=-{\partial_t\vec B},
\end{equation}
where $\vec B$ is the magnetic field (we {assume permeability $\mu=1$}). The magnetic field in \req{MEq} modifies the expression for the energy current from
\req{je} to
\begin{equation}\label{jep}
\vec j_{\varepsilon}=-\kappa T\vec{\nabla} S+\vec{\mathcal P};
\qquad  4\pi\vec{\mathcal P}\equiv c\vec{E}\times \vec{B},
\end{equation}
where  $\vec{\mathcal P}$ is the Poynting vector acting as an external source for the energy dynamics inside the active layer.

We consider ``probe'' radiation incident at an angle $\theta$ and distinguish two polarizations: when electric field $\delta E \parallel \hat y$ (transverse electric or TE polarization) or when $\delta E$ has a component along $z$ (transverse magnetic or TM polarization, see Fig. 2d). Symmetry dictates that electric fields associated with TE radiation cannot interact with the charge oscillations of the longitudinal mode Eq. \eqref{adiab} so that no significant changes in $R(\omega)$ may occur (the other way to see this is to notice that the Poynting vector is $\vec{\mathcal P}\parallel\hat y$ but the only important spatial variation is along $x$ so $\vec{\nabla}\cdot\vec{\mathcal P}=0$). The absence of a pump dependent correction for TE polarization is a key qualitative result of our model.

For TM polarization the Poynting vector of the incident wave indeed acts as a source in the energy conservation Eq. \eqref{dte}, modifying Eq. \eqref{L} to
\begin{equation}\label{Ls}
 4\pi \hat L\begin{pmatrix}\delta \zeta;
   \ \delta\varepsilon;\ T\delta S\end{pmatrix}^T=c(\partial_xB)E^{*}
 \begin{pmatrix}0;\ 1;\ 0\end{pmatrix}^T,
\end{equation}
where the condition $\omega l_0/c\ll1$ implies that the dependence of $B$ on $z$ can be neglected. The linear perturbation of the layer is maximal when the frequency of the probe is close to the real part of the frequency $\omega_0$ [Eq. \eqref{w0}].

We now calculate the frequency dependent reflectivity $R=|r|^2$ in terms of the amplitude, $r(\theta, \omega)$, of the reflected portion of a TM wave incident at angle $\theta$:
\begin{equation}\label{rw}
r(\theta,\omega)=\left(4\pi\cos\theta-c\tilde Z\right)\Big/\left(4\pi\cos\theta+c\tilde Z\right).
\end{equation}

The total impedance  $\tilde Z(\theta,\omega)$ is defined as
\begin{equation}\label{Z1}
\frac1{\tilde Z(\theta,\omega)}\equiv
\frac c{4\pi}\frac{B_y(z=0)}{E_x(z=0)}=\frac{\int_0^{\infty}j_D^xdz}{E_x(z=0)},
\end{equation}
where the last equation is obtained by integration of the first Maxwell equation \eqref{MEq} over $z$ within the sample.

The total displacement current is given by
\begin{equation}\label{jd}
\int_0^{\infty}j_D^xdz=\left[{Z_0(\theta,\omega)}\right]^{-1}{E_x(z=l_0)}+\int_0^{l_0}j_D^xdz,
\end{equation}
where $Z_0(\theta,\omega)$ is the equilibrium impedance in the bulk and the second term is always small for $\omega l_0/c\ll1$. It is the field $E_x$ that drastically changes across the active layer; in fact, for $\omega l_0/c\ll1$, $\vec{\nabla}\times\vec E\approx0$ and we obtain
\begin{equation}\label{Ex}
E_x|_{z=l_0}-E_x|_{z=0}=\int_0^{l_0}\partial_x\delta E_zdz=\int_0^{l_0}\frac{\partial_x\delta D_z}{\epsilon_r}dz.
\end{equation}

Finding $\delta D_z$ from Eqs. \eqref{dE} and \eqref{Ls} we obtain the angular dependence of the non-equilibrium impedance
\begin{equation}\label{Z}
\tilde Z=Z_0(\theta,\omega)+ Z_\textrm{{neq}}(\theta,\omega);\quad\,
Z_\textrm{{neq}}\equiv \sin^2\theta\,Y(\omega),
\end{equation}
where $Z_0(\theta,\omega)$ has to be extracted from equilibrium experimental measurements \footnote{For K$_3$C$_{60}$ we computed the impedance $Z_0$ using $\epsilon_r=1.6$, $\sigma_1(\omega)/\sigma_0=0.12+0.8\sqrt{4\log^2(\omega/10)+0.1^2}$ for $\omega<10$ and $\sigma_1(\omega)/\sigma_0=0.2$ for $\omega>10$, $\sigma_2(\omega)/\sigma_0=2/3-1.98\log^2(\omega/3)$ for $\omega<8$ and $\sigma_2(\omega)/\sigma_0=1/30-1.98\log^2(\omega/20)$ for $\omega>8$, where $\sigma_0=400\um{\Omega^{-1}cm^{-1}}$ and $\omega$ is expressed in $\um{meV}$. This choice of $\sigma$ reproduces the important low frequency features of the reflectivity in Fig. 2a of Ref. \cite{K3:C60}. For LBCO we reproduced the equilibrium data from Fig. 2b.2 of Ref. \cite{LBCO1} using $\sigma=3.5\um{\Omega^{-1}cm^{-1}}$ and $\epsilon_r=50$.}.

The factorization of the non-equilibrium contribution $Z_\textrm{{neq}}$   into angle and  frequency dependent terms
is a distinctive feature of the active layer model. The specific  $\theta$-dependence shown in  Eq. ~\rref{Z} is a consequence of the assumed domain shape.  A more complex domain structure would produce a more complicated $\theta$-dependence.

The function $Y(\omega)$ is formally expressed as
\begin{equation}\label{Y1}
\begin{split}
\frac{c }{4\pi}Y(\omega)&=\frac1{\epsilon_r}\frac{\omega^2}{ c}\int_0^{l_0}dz_1dz_2
\frac1{E^{*}(z_1)}
\mathcal{L}_{z_1,z_2}E^{*}(z_2);\\
\mathcal{L}&=\left[\left(-{G}/{I},1,-1\right)\hat L^{-1}\begin{pmatrix}0;\ 1;\ 0\end{pmatrix}^T\right].
\end{split}
\end{equation}
We neglect the factor $G/I$ which is of the order of the rate at
which the state relaxes back to equilibrium divided by the frequency:
$\frac GI\sim\frac{\partial I/\partial \zeta}{\omega}\ll1$. Explicit
calculation within the model leading to Eq. \eqref{w0} gives
\begin{equation}\label{Y2}
\frac{c}{4\pi}Y(\omega)=-\frac{\omega_El_0}{\epsilon_r c}\left(1-i\frac{\gamma}{\omega}\right)\left(\frac{\tan[k(\omega)l_0]}{k(\omega)l_0}-1\right),
\end{equation}
where $k(\omega)$ is found from Eq. \eqref{k2}. The function $Y(\omega)$ (see Fig. 3a) vanishes as $\omega\rightarrow 0$. For small $\omega$, $\mathrm{Re}(Y)<0$ and $Y$ has poles at the eigenfrequencies given in Eq. \eqref{w0}. The high frequency behavior cannot be obtained from Eq. \eqref{Y2},
valid only for $\omega<\frac c{l_0}$. The remarkable feature of Eq. \eqref{Y2} is that the pre-factor $(\omega_El_0/c)$ can easily exceed unity;
the origin of this largeness is the sensitivity of the entropy production to the integral of motion $\varepsilon$. 
\begin{figure}[t]\label{1e}
\centering
\includegraphics[width=0.47\columnwidth]{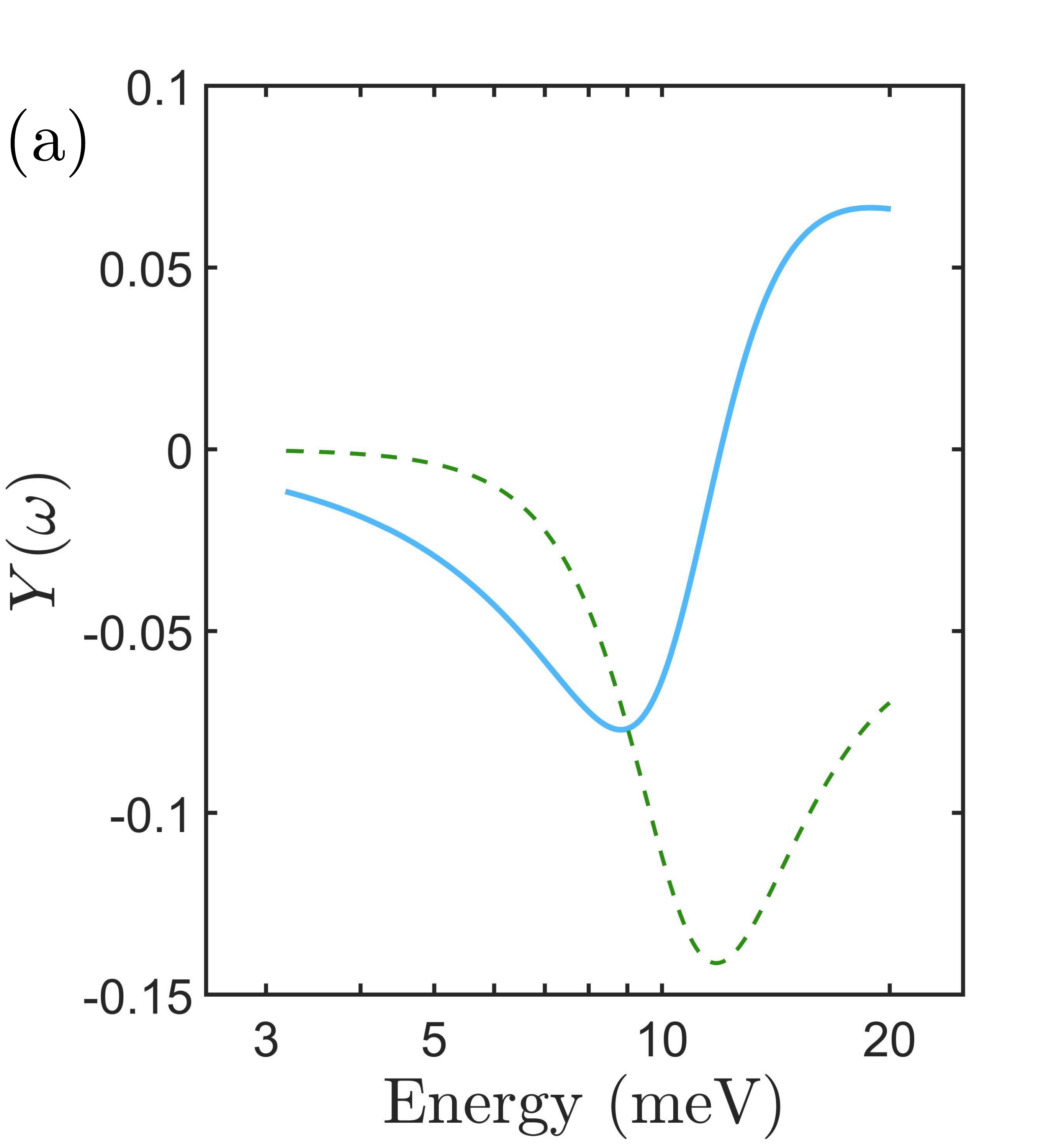}\,\,\,\,
\includegraphics[width=0.47\columnwidth]{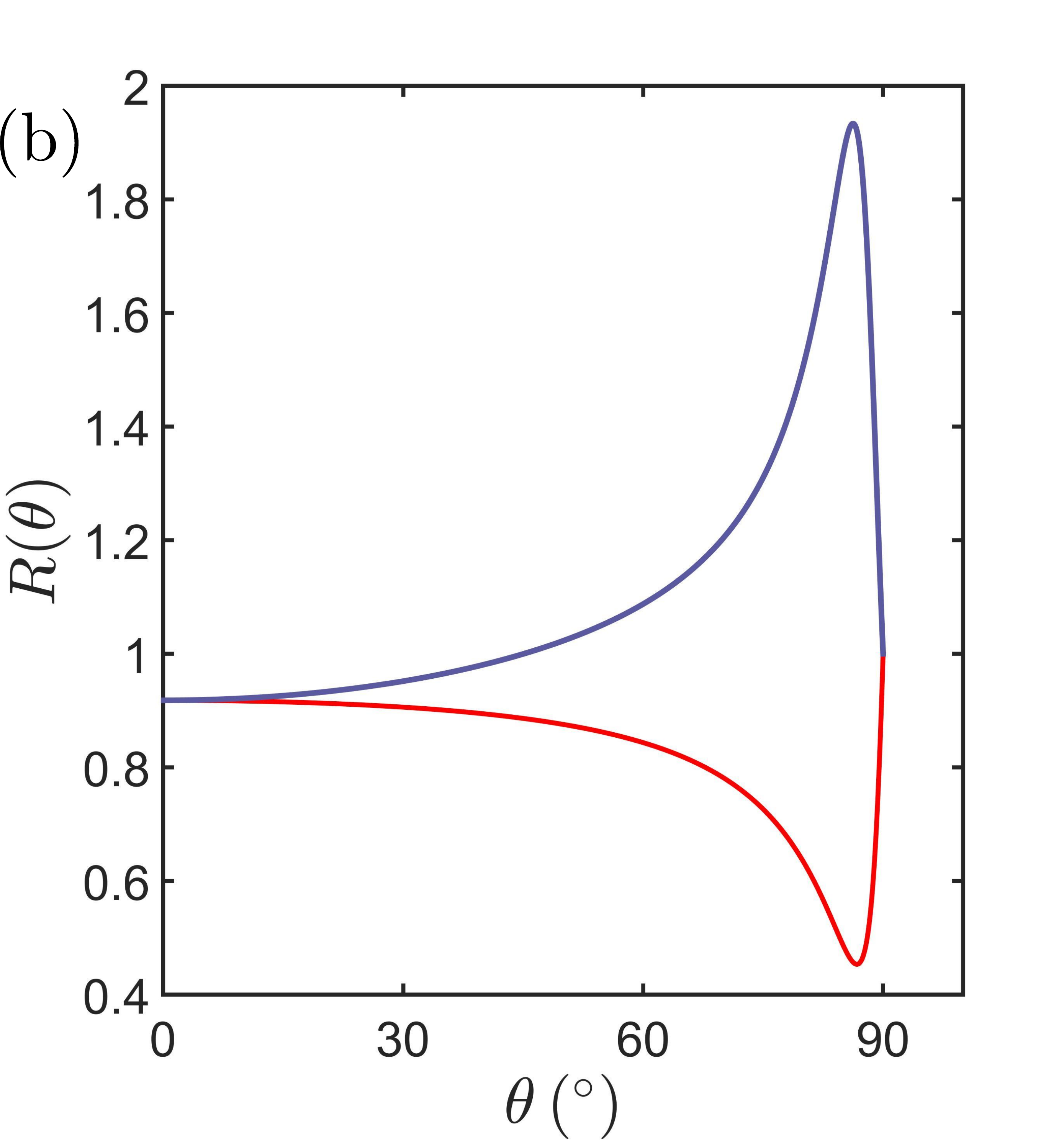}
\caption{\scriptsize{(Color online) (a) Plot of the real (solid line) and imaginary (dashed line) part of $Y(\omega)$. (b) Theoretical $R(\theta)$ of K$_3$C$_{60}$ at $\omega=6\um{meV}=1.44\um{THz}$ in equilibrium (red) and non-equilibrium (blue); notice the marked dependence on the angle and the presence of a region where $R>1$.
Plots are for the parameters of Fig. 1a.}}
\end{figure}

We used Eqs. \eqref{rw}, \eqref{Z}, \eqref{Y2} to calculate the reflectivity. From Eq. \eqref{Z} we see that at $\theta=0$ (normal incidence) the non-equilibrium effects are not visible in $R$, while from Eq. \eqref{rw} we see that at $\theta=\pi/2$, $|r|=1$ for both equilibrium and non-equilibrium states. For $0<\theta<\pi/2$, non-equilibrium effects are evident in the reflectivity, see Fig. 3b. Intuition may be gained considering the Hagen-Rubens limit $|\tilde Z|\ll1$ \cite{Hagen1903} in which $R(\omega)\approx 1-4\mathrm{Re}[Z_0+\sin^2\theta Y(\omega)]/\cos\theta$, showing that the reflectivity is enhanced relative to equilibrium for $Y<0$ and suppressed for $Y>0$. The large value of $\omega_El_0/c$ means that $\mathrm{Re}(Z)$ can become \textit{negative} for $\omega\lesssim\mathrm{Re}(\omega_0)$, leading to $R>1$; such amplification is allowed in a non-equilibrium system. Notice however that there is no spontaneous emission instability (no lasing).

We briefly mention nonlinear response effects implied by our model:
(i) Second harmonic generation (SHG) is made possible by the non-zero value of the spontaneous polarization $E^{*}$, which reduces  the symmetry to uniaxial. The SHG signal is maximal for excitation frequencies near $\omega_0/2$ and $\omega_0$, corresponding to resonances in the outgoing or incoming state respectively.
(ii) A parametric resonance instability may lead to radiation at frequencies $\sim\omega_0$ in response to an incident wave of frequency close to $2\omega_0$
\cite{*[{See e.g. $\ddagger$ 27, 28 of }][]
LLMechanics}; the observable features are the same as those of the recently discussed  ``Floquet time crystal'' state \cite{Nayak2016}.

In conclusion, we plotted $R(\omega)=|r(\omega,\theta=45^{\circ})|^2$ for ``sensible'' parameters values in Figs. 1a and 1b. The resemblance with the experimental data is very reasonable, even though we cannot make any definite conclusion until the data on polarization and angular dependence are available.

\begin{acknowledgments}
We are grateful to A. Andreev, D. Basov, A. Cavalleri, M. Dyakonov, Y. Galperin, A. Georges, and L. Glazman for helpful discussions of the results of this paper. Support was provided by the Basic Energy Sciences Division of the Office of Science of the United States Department of Energy under Grant No. DE-SC0018218 (A.M. and G.C.) and by the Simons Foundation (I.A.).
\end{acknowledgments}

\bibliographystyle{apsrev4-1}
\bibliography{bibliography}

\end{document}